# SIMULATION OF COLOR BLINDNESS AND A PROPOSAL FOR USING GOOGLE GLASS AS COLOR-CORRECTING TOOL

H.M. de Oliveira* (qPGOM), J. Ranhel*, R.B.A. Alves**

* Biomimicry, Neurocomputational Intelligence, Cognitive Agents (BINAC), Recife, Brazil
** Center of Informatics (CIn), Recife, Brazil

e-mail: hmo@ufpe.br at www.binac.org

*Abstract:* The human visual color response is driven by specialized cells called cones, which exist in three types, viz. R, G, and B. Software is developed to simulate how color images are displayed for different types of color blindness. Specified the default color deficiency associated with a user, it generates a preview of the rainbow (in the visible range, from red to violet) and shows up, side by side with a colorful image provided as input, the display correspondent colorblind. The idea is to provide an image processing after image acquisition to enable a better perception of colors by the color blind. Examples of pseudo-correction are shown for the case of Protanopia (red blindness). The system is adapted into a screen of an i-pad or a cellphone in which the colorblind observe the camera, the image processed with color detail previously imperceptible by his naked eye. As prospecting, wearable computer glasses could be manufactured to provide a corrected image playback. The approach can also provide augmented reality for human vision by adding the UV or IR responses as a new feature of Google Glass.

**Introduction**

How people with color blindness see the pictures? Color vision deficiency is the inability to perceive color differences under ordinary lighting conditions. It is recognized that color blindness affects a non-negligible proportion of the population [1], [2]. Already in the seventeenth century it was discovered that the retina was responsible for detecting light, not the cornea. Johannes Kepler and René Descartes did the major advances of the time.... Kepler proposed that the image was focused on the retina. A few decades later, Descartes showed that Kepler was correct. For the demonstration, he surgically removed the eye of an ox [3]. He also postulated that the image was inverted, and conclude that astigmatism is the result of improper curvature of the cornea. John Dalton published in 1798 "Extraordinary facts relating to the vision of colours" [4] after the realization of his own color blindness. In the 1830s, several German researchers used a microscope to examine the retina. Two types of grid cells were found in the retina - rods and cones - so called because of their shape when viewed in the microscope. Max Schultze (1825-1874) found that the cones are the color receptors in the eye and the rods are not color sensitive, but very sensitive to low light levels. In the human eye there are plenty more rods than cones. Roughly there are $10^9$ rods/retina and $10^6$ cones/retina [5].

During the nineteenth century, visual pigments in the retina were discovered. Dissecting the eyes of frogs, it was found that the color be modified with incident light (the retina is photosensitive). The membrane discs located in the rods contain rhodopsin (Mr 40,000) a light-sensitive protein. In cones, there are different types of rhodopsin, which allows color vision. The sensitivity is due to a pigment called rhodopsin. Later studies showed that this is a protein and the Opsin was also discovered [6]. Pigments are also found in the cones and there are three types of cone cells, according to the pigment they contain. An original theory for color vision was proposed by Thomas Young (1773-1829) circa 1790, even before the discovery of cone cells in the retina [7], [8]. Young pioneered to propose that the human eye sees only the three primary colors. The absorption spectra of the three systems overlap, and combine to cover the visible spectrum. Despite how simplistic theory, the work of Young established the foundations of the theory of color vision (the Young-Helmoltz theory, in honor of the two early and late 19th century figures who played the substantial role in putting this scheme forward). In humans, there are usually three types of cones, whose spectral sensitivities differ; conferring a trichromatic vision. In mammals, there are cones with sensitivity (max) at wavelengths 424 nm, 534 nm and 564 nm (Fig.1).

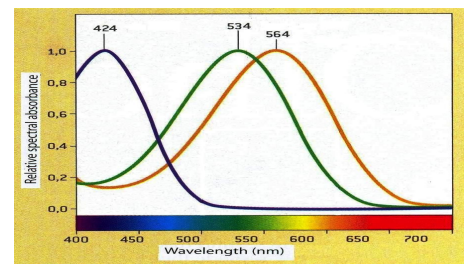

Figure 1: Cone response on the human eye as a function of wavelength of the incident light: spectral absorption of S, M and L purified rhodopsin (Source [12]).

In the human eye, the cones are maximally receptive three wavelength bands - short, medium and long cones called S, M, L. (Table 1), but they are also often referred to as blue cones, green cones, and red cones, respectively [8]. The "blue" opsin gene is located on Chromosome 7, while the "red" and "green" genes are located on Chromosome X [9]. The color blindness, such as the inability to distinguish red and green, is commonly inherited. If some genes that produce photopigments are missing or damaged, color blindness will be expressed in males with a higher probability than in females because males only have one X chromosome [10]. Primates and other mammals are dichromatic and others just do not even have color vision. In birds and other creatures of the animal kingdom have definite evidence that ultraviolet light plays a key role in the perception of color images [11]. They have tetrachromatic view with four different types of cones. The birds, lizards, turtles, among others, are capable of higher mammals' vision. "Why do men should not improve their vision?" The spectral sensitivities of the cones differ; one is most sensitive to short wavelengths, one to medium wavelengths, and the third to medium-to-long wavelengths within the visible spectrum, with their peak sensitivities in the blue, green, and yellow-green regions of the spectrum [12]. However, the link between visual perception and physiology is rather intricate. Details on colors and their significance, as well as many facts concerning the influence of color on the perception of the world are discussed in [13].

Table 1. Common cones existent in a human eye

| Cone | pigment | spectral response | Peak |
|---|---|---|---|
| S | β (blue) | 400 - 500 nm | 424 nm |
| M | γ (greenish blue) | 450 - 630 nm | 534 nm |
| L | ρ (yellowish green) | 500 - 700 nm | 564 nm |

**Abnormal Vision (color blindness)**

There are three types of congenital color blindness: Monochromacy is the lack of ability to distinguish colors caused by cone defect or absence. Dichromacy is a moderately severe color vision defect in which one of the three basic color mechanisms is absent or not functioning. Anomalous trichromacy is when one of the three cone pigments is altered in its spectral sensitivity (impairment, rather than loss, of trichromacy). The pigments in the L and M cones are encoded on the X chromosome of the genome, and its deficiency leads to the most common forms of blindness in color. The frequencies of light stimulate each among these types of receptors with varying intensity. There are several types of mild disability, depending on the mutation in Opsin [14]. Sometimes there is no answer on a given spectral band, leading to a dichromatic vision (red- or green-dichromatic. These are known as Protanopia and Deuteranopia, respectively). In most cases, photoreceptors for R and G do exist, but the change in protein leads to a change in the spectral range absorbed, resulting in abnormal vision (Protanomaly: *red-anomalous trichromatic*; Deuteranomaly: *green-anomalous trichromatic*). A visual example of protanopia, tritanopia and deuteranomaly is shown in Fig.2b-d, respectively, for the image corresponding to the rainbow spectrum shown in Fig.2a.

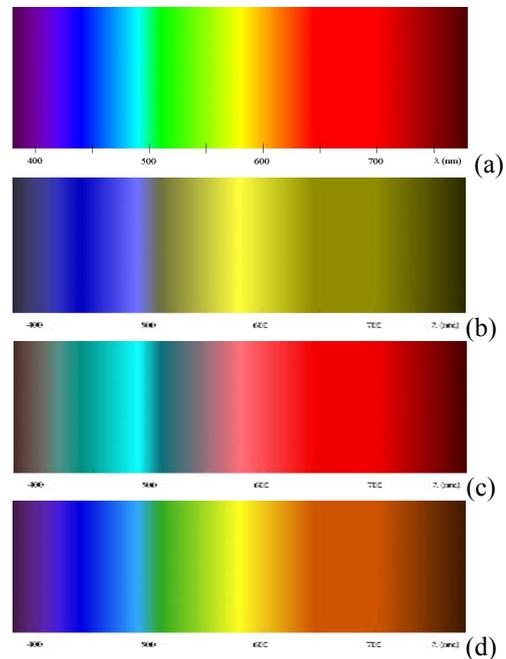

Figure 2: Straightforward examples of abnormal vision for the original spectrum: a) original rainbow, b) view of colorblind with protanopia, c) view of colorblind with tritanopia, d) view of colorblind with deuteranomaly.

**Visual Simulation of Color Blindness**

Freeware software was developed to provide a view how colorblind people perceive a color image and several kind of color blindness are considered [15]. The absence of L cones yields a misleading perception of colors towards the red (Fig. 2b). The Ishihara Color Test is a "color perception test" devised in 1917 by Shinobu Ishihara 石原 忍, to screen military recruits for abnormalities of color vision [16]. A guileless example of pseudo-correction is shown for the case of Protanopia, by considering images presented in Fig.3. These plates have been presented to three people carrying protanomaly. No one was able to identify the numbers displayed in Fig.3 [this can be reproduced at http://www2.ee.ufpe.br/codec/colorblindnesssimulator.html].

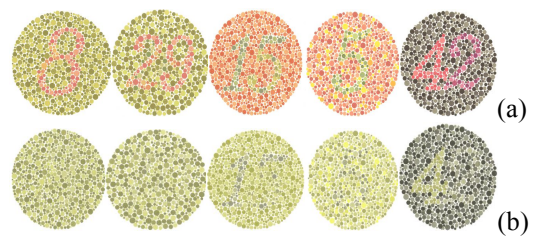

Figure 3: Ishihara Color Test (plates 2, 3, 6 and plate 10) is shown for a person with Protanopia blindness. (a)

Original plates, (b) Images as seen by the color blindness (images can be retrieved from [15]).

**A case study: unsophisticated protanomaly "correction"**

An interesting analysis of color blindness through different decoding models (RGB, fuzzy 3-8, HSV) is addressed in [18]. The issue may not be a color correction, but to provide some "information" in the captured image so imperceptible color differences can now be recognized by the colorblind. Therefore, this is not necessarily about color correction itself, but to provide some "change" in the image automatically captured so imperceptible color differences can now be recognized by the colorblind. People that are not able to visualize red (a null response as in Fig.3b) may have the red component of the image replaced by a grayscale. As a result, this approach applied to the images of Fig.3 yield those shown is Fig.4, where undetected numbers (Fig.3b) by the naked eye of the colorblind can now be identified. In fact, all people under test were promptly able to identify such numbers in the monochromatic version of Fig.4b (derived from the red-component of the corresponding Fig.3a).

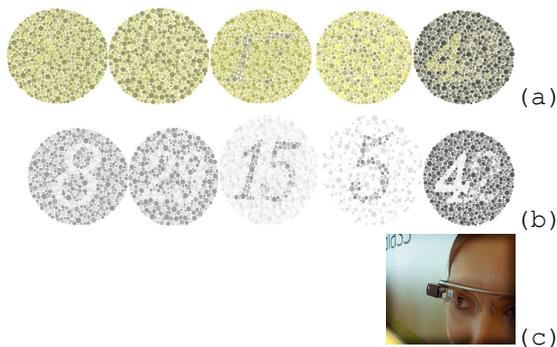

Figure 4: Protanopia blindness. (a) Original color image shown in Fig.3a as seen by the protanomaly colorblind. (b) red-luminance after processed, displaying the red monochromatic version (c) Such images could be viewed with aid of Google glass.

A further issue is providing an image correction for individuals with some abnormality of color perception. In these cases it is possible to design some sort of equalization-type correction [17]. An auxiliary image can be exhibited with increased luminance so as to make amends on the response of the colorblind.

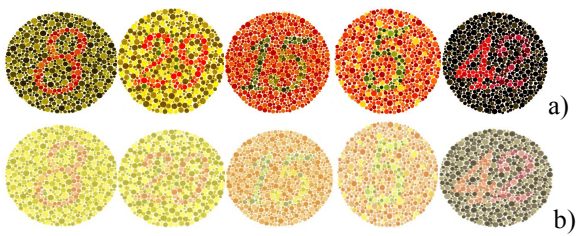

Figure 5: Protanomaly blindness correction: a) color plates of Fig.3a after processing. b) "corrected" images as seen by the protanomaly colorblind. Compare the picture seen in b with those of Fig.3b, without any correction.

Therefore, a choice to enhance visualization of color blinders is to generate a "helper image" through color adjust by setting the saturation level set as -100. The results for the plates shown in Figure 4 are shown in Figure 6. Some recent approaches for "removing" color blindness using image processing can also be found in [19], [20]. The process of adjusting the level of color saturation at -100 offered here is much simpler than the latter techniques.

**Discussion**

There are a couple of devices for helping colorblind people, for example, one can purchase the "*Oxy-Iso Colorblindness Correction Medical Glasses*" by a couple of hundred dollars on the internet. Such glasses act as passive filters for green color (M cones) so that red and blue colors pass through the lens (Fig.6). It explains the magenta color of the lens. Although the device can help, a possible collateral damage is that users see a 'pink world'.

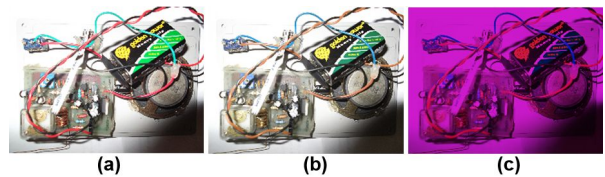

Figure 6: (a) an electronic circuit board original image, (b) anomalous color blind view, and (c) color blind 'correction' as seen through an Oxy-Iso glass.

On the other hand, nowadays, electronic devices are noticeable source of images, from computers screen to TVs, from tablets to mobiles. Modifying electronic signal intensities is quite simple, be it analogically or by digital processes. Once electronic image processing is ubiquitous, we investigate how to introduce image modifiers that help people to improve 'accessibility' when looking the world mediated by electronic devices.

We have worked on filters that perform modification in the same manner the referred glasses do. However, in addition to passive filters we have investigated active functions, such as a blinking function over the image areas under modification before applying the effects permanently. It may call user's attention over where modifications take place. Another bet is to enhance object borders presenting certain colors. These and other functions are yet under investigation and results are preliminary. Nevertheless, we understand that active modifiers may be more efficient, besides easily implemented in electronics. User's personal preferences will decide which functions are more functional and deserve more attention. Therefore, we propose that electronic media may be an efficient, adaptable, ubiquitous, and practical way for helping colorblind people to see details they cannot. We investigate simple and inexpensive circuits that, added to electronic image generators, would allow more accessibility to blindness persons.

**Conclusions**

Freeware software was developed with html interface that provide a simple way to view how colorblind people perceive a loaded color image. Several kind of color blindness is available. A naive image processing is offered as a way to improve the image perception so former imperceptible details can now be recognized by a colorblind. This software was adapted to i-pad (or possible to an android cell phone), in which the colorblind observe the camera, the image processed with color detail previously imperceptible by his naked eye.

As prospecting, glasses could be manufactured with embedded correction algorithm with an undersized external camera to image capture, and Liquid crystal on silicon (LCoS) or RGB-LED screen on the inner face of the lens to the corrected image playback. It could also be popped in as a new Google Glass feature. This can be useful, very effective and comfortable for people with colorblindness. Another interesting issue is to enhance the visual capability of humans by introducing *a tetrachromatic vision* [21] or even *pentachromatic vision* (augmented reality). The option to add the UV response (e.g. 370 nm peak) and/or IR response (e.g. 630 nm peak) in the synthesized image can be enabled or not.

**Acknowledgements**

The authors thank the three colorblind colleagues (A.J.P.A, G.G.S., W.A.C.) who kindly proposed to cooperate with this investigation according to written permission. BINAC thanks to PROPESQ / UFPE for partial financial support to this project.

**References**

[1] M. Ananya, Color Blindness Prevalence. Health. http://www.news-medical.net/health/Color-Blindness-Prevalence.aspx Retrieved 12 April 2014.

[2] R.H. Post, Population differences in red and green color vision deficiency: A review, and a query on selection relaxation. *Biodemography and Social Biology,* vol.29, pp.299-315, 1982.

[3] R. Descartes. Discourse on Method, Optics, Geometry, and Meteorology, Hackett Publishing, 2001.

[4] J. Dalton, Extraordinary facts relating to the vision of colours: with observations. Cadell and Davies, London, 1798.

[5] B. Alberts et al. Essential Cell Biology, Garland Science, 2013.

[6] C.U.M Smith, Biology of Sensory Systems. John Wiley & Sons, Ltd, 2000.

[7] P.K. Kaiser, R.M. Boynton, Human Color Vision, Optical Society of America, 1996.

[8] E.H. Land, Color Vision and the Natural Image, Part I. *Proceedings of the National Academy of Sciences of the United States of America*, vol.45, p.115, 1959.

[9] J. Nathans, D. Thomas, D.S. Hogness, Molecular Genetics of Human Color Vision: the genes encoding blue, green, and red pigments. *Science*, vol.232, pp.193-202, 1986.

[10] N.R. Carlson, Physiology of Behavior, 11th Edition. Pearson, 2012.

[11] L.S. Carvalho, J.A. Cowling et al. The Molecular Evolution of Avian Ultraviolet- and Violet-sensitive Visual Pigments, *Molecular Biology and Evolution,* vol.24, pp.1843-1852, 2007.

[12] J.L. Schapf, T.W. Kraft, D.A. Baylor, Spectral Sensibility of Human Cone Photoreceptors. *Nature*, vol. 325, pp.439-441, 1987.

[13] D. Purves, R.Beau Lotto, Why We See What We Do: An Empirical Theory of Vision, Sinauer Ass. Inc. Pub, MA USA, 2003.

[14] J. Nathans, D. Thomas, D.S. Hogness, Molecular Genetics of Human Color Vision: the genes encoding blue, green, and red pigments. *Science*, vol.232, pp.193-202, 1986.

[15] H.M. de Oliveira, R.B.A. Alves, http://www2.ee.ufpe.br/codec/colorblindnesssimulator.html Retrieved April 2014.

[16] S. Ishihara, Tests for Color-blindness, Handaya, Tokyo, Hongo Harukicho, 1917.

[17] G. Ballou, Handbook for Sound Engineers. Focal Press, 2013.

[18] C. Lu, Illustrating Color Evolution and Color Blindness by the Decoding Model of Color Vision:
http://arxiv.org/ftp/arxiv/papers/1101/1101.2243.pdf

[19] R. Kulshrestha, R.K. Bairwa, Review of Color Blindness Removal Methods using Image Processing, *Int. J. of Recent Research and Review*, vol.6, pp.18-21, June 2013.

[20] R. Kulshrestha, R.K. Bairwa, Removal of Color Blindness using Threshold and Masking, *Int. J. Adv. Res. in Computer Sci. and Soft. Eng.*, Vol.3, June pp.218-221, 2013.

[21] T.H. Goldsmith, What Birds See, *Scientific American*, vol.295, pp.68-75, 2006.